\documentstyle[twoside,fleqn,espcrc2,epsf]{article}

\newcommand{\AmS}{{\protect\the\textfont2
  A\kern-.1667em\lower.5ex\hbox{M}\kern-.125emS}}

\title{
\vspace*{-35pt}
{\normalsize \hfill {\sf UTCCP-P-51}} \\
\vspace*{-6pt}
{\normalsize \hfill {\sf Sept.\ 1998}} \\
The static quark potential in full QCD
\thanks{talk presented by T.~Kaneko
at Lattice98, Boulder, Colorado, USA, 13--18 July 1998.
}}

\author{CP-PACS Collaboration :\\
        S.~Aoki\rlap,\address{Institute of Physics, University of
        Tsukuba, Tsukuba, Ibaraki 305-8571, Japan}
        G.~Boyd\rlap,\address{Center for Computational Physics,
        University of Tsukuba, Tsukuba, Ibaraki 305-8577, Japan}
        R.~Burkhalter\rlap,$^{\rm a,b}$
        S.~Ejiri\rlap,$^{\rm b}$
        M.~Fukugita\rlap,\address{Institute for Cosmic Ray Research,
        University of Tokyo, Tanashi, Tokyo 188-8502, Japan}
        S.~Hashimoto\rlap,\address{High Energy Accelerator Research Organization
        (KEK), Tsukuba, Ibaraki 305-0801, Japan}
        Y.~Iwasaki\rlap,$^{\rm a,b}$
        K.~Kanaya\rlap,$^{\rm a,b}$
        T.~Kaneko\rlap,$^{\rm b}$
        Y.~Kuramashi\rlap,$^{\rm d}$
        K.~Nagai\rlap,$^{\rm b}$
        M.~Okawa\rlap,$^{\rm d}$
        H.~P.~Shanahan\rlap,$^{\rm b}$
        A.~Ukawa\rlap,$^{\rm a,b}$ and
        T.~Yoshi\'e$^{\rm a,b}$ }

\begin{document}

\begin{abstract}
We report results on the static quark potential in two-flavor full QCD.
The calculation is performed for three values of lattice spacing 
$a^{-1}\approx 0.9$, $1.3$ and $2.5$ GeV 
on $12^3{\times}24$, $16^3{\times}32$ and $24^3{\times}48$
lattices respectively, 
at sea quark masses corresponding to $m_\pi/m_\rho \approx 0.8$--$0.6$. 
An RG-improved gauge action and a tadpole-improved SW clover quark action 
are employed. We discuss scaling of $m_{\rho}/\sqrt{\sigma}$ and 
effects of dynamical quarks on the potential.
\end{abstract}

\maketitle
\setcounter{footnote}{0}
\section{Introduction}

Interest in the static quark potential calculated in full QCD 
is multi-fold, the foremost being the expectation that a flattening of 
the potential due to quark pair creation would be 
observed at large separations.
An increase in the Coulomb coefficient of the potential at small 
separations is another effect expected from sea quarks. 
A number of studies have recently been reported exploring these 
aspects of the full QCD static potential
\cite{Lat98.Kuti,stringbreaking.Lat97}.  

Recently the CP-PACS Collaboration has started a systematic effort 
toward a full QCD simulation with two flavors of dynamical 
quarks\cite{Lat98.Ruedi}. 
In this article we present results on the static potential calculated 
as a part of this program. 

\vspace{-1pt}
\section{Simulations and measurements}
\label{sec:simulation}

The CP-PACS simulation of full QCD is carried out 
for three values of lattice spacing in the range 
$a^{-1} \approx 0.9$--$2.5$~GeV 
and four different sea quark masses corresponding to 
$m_\pi/m_\rho \approx 0.8$--$0.6$.  An RG-improved 
gauge action and a tadpole-improved SW clover action are employed 
to reduce discretization errors.  
Light hadron masses 
are calculated at every fifth
trajectories
\cite{Lat98.Kanaya}, 
while the static potential calculation 
uses a subset of the configurations
as listed in Table~\ref{tab:param}
where other relevant run parameters are also given.

\begin{table}[t]
\setlength{\tabcolsep}{0.2pc}
\caption{Parameters of potential measurements.
Statistics for $m_{SL}$ is given in the square brackets.
Lattice scale is from $m_\rho\!=\!0.768$~GeV at the physical point.}
\label{tab:param}
\begin{tabular}{llll}
\hline
 lattice   & $K_{\rm sea}$ & conf & $m_\pi/m_\rho$ \\
\hline
$12^3{\times}24$  &  0.1409   &  985 &   0.806(1) \\ 
$\beta\!=\!1.80$     &  0.1430   &  845 &   0.753(1) \\
$c_{SW}\!=\!1.60$     &  0.1445   & 1063 &   0.696(2) \\
$a^{-1}_{\rho}\!=\!0.917(10)$~GeV    &  0.1464   &  680 &   0.548(4) \\
\hline
$16^3{\times}32$  &  0.1375   &  659 [375] &   0.805(1) \\ 
$\beta\!=\!1.95$    &  0.1390   &  690 [199] &   0.751(1) \\
$c_{SW}\!=\!1.53$     &  0.1400   &  659 [169] &   0.688(1) \\
$a^{-1}_{\rho}\!=\!1.288(15)$~GeV    &  0.1410   &  492 [169] &   0.586(3) \\
                  \cline{2-4}
\hline
$24^3{\times}48$  &  0.1351   &  250 &   0.800(2) \\ 
$\beta\!=\!2.20$    &  0.1358   &  269 &   0.752(3) \\
$c_{SW}\!=\!1.44$     &  0.1363   &  322 &   0.702(3) \\
$a^{-1}_{\rho}\!=\!2.45(9)$~GeV    &  0.1368   &  253 &   0.637(6) \\
                  \cline{2-4}
\hline
\end{tabular}
\vspace{-25pt}
\end{table}

We extract the static potential from smeared 
Wilson loops by an exponential fit
over the range $T \approx 0.4$--$0.8$~fm, above which
noise quickly dominates over the signal.
The string tension $\sigma$ 
and the Coulomb coefficient $\alpha$
are determined by fitting the potential result
to a form
$V(R) = V_0 - \alpha / R + \sigma \cdot R$.
Errors are estimated by the jack-knife method
with a bin size corresponding to 50 HMC trajectories.

We also carry out a measurement of the potential 
with 100 configurations in quenched QCD at $\beta\!=\!2.7$ 
to compare with 
full QCD results obtained at $\beta\!=\!2.2$.

At $\beta\!=\!1.95$, we calculate the static-light 
meson propagator using an exponential smearing function for the 
source.
The mass of the static-light meson
is extracted by 
making fits over an interval  $T \approx 0.3$--$1.1$~fm. 

\vspace{-1pt}
\section{Scaling of $m_{\rho}/{\protect \sqrt{\sigma}}$ }

\begin{figure}[tb]
\vspace{-4mm}
\begin{center}
\leavevmode
\epsfxsize=7.5cm
\epsfbox{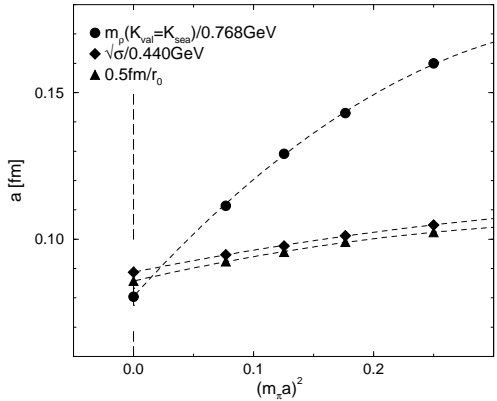}
\end{center}
\vspace{-18mm}
\caption{
The lattice spacings at $\beta\!=\!2.2$ determined from 
$m_{\rho}$, $\sigma$ and 
$r_0$ as a function of $m_{\pi}^2a^2$ together with 
quadratic chiral extrapolations.}
\label{fig:scale}
\vspace{-19pt}
\end{figure}

In Fig.~\ref{fig:scale}
the lattice spacings determined 
from $m_\rho$, the string tension $\sigma$ (where we employ the value 
$\sqrt{\sigma}\!=\!0.44$ GeV) and the Sommer scale $r_0$
at each sea quark mass on the $24^3{\times}48$ lattice
are plotted as a function of $m_\pi^2a^2$.
The values extrapolated quadratically in $m_\pi^2a^2$ to the chiral limit
are also shown. 
We observe that the three lattice spacings
converge within 10\% in the chiral limit. 
The consistency of scale observed here at $a\approx 0.08$~fm holds also 
at larger lattice spacings. 

\begin{figure}[tb]
\vspace{-4mm}
\begin{center}
\leavevmode
\epsfxsize=7.5cm
\epsfbox{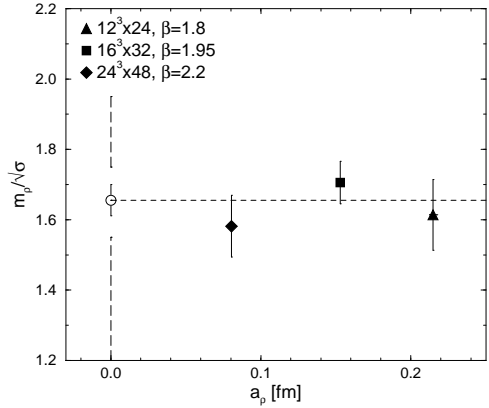}
\end{center}
\vspace{-18mm}
\caption{
Ratio 
$m_{\rho}/{\protect \sqrt{\sigma}}$ 
in the chiral limit  as a function of $a_{\rho}$ 
determined from $m_\rho$. An estimate of the continuum value from a 
constant fit is also shown. }
\label{fig:ratio}
\vspace{-19pt}
\end{figure}

In Fig.~\ref{fig:ratio} we plot the ratio 
$m_\rho/\sqrt{\sigma}$ in the chiral limit 
as a function of $a_\rho$ fixed by $m_\rho$.
With our choice of the clover coefficient for the quark action,
the leading $O(g^2 a)$ scaling violation 
is expected to be small\cite{csw}.
The flat behavior of the ratio supports this expectation,
and also suggests that 
the remaining higher order contributions are small as well.
We then adopt a constant fit, and find $m_\rho/\sqrt{\sigma}\!=\!1.66(4)$ 
as an estimate of the continuum value.  
If we employ $m_\rho\!=\!0.768$~GeV, this leads to 
$\sqrt{\sigma}\!=\!0.463(11)$~GeV, which is now a prediction of our full QCD
simulation. 
This value is about 5\% larger than 0.44 GeV used in Fig.~\ref{fig:scale}.

\section{String breaking effect}

\begin{figure}[tb]
\begin{center}
\vspace{-4mm}
\leavevmode
\epsfxsize=7.5cm
\epsfbox{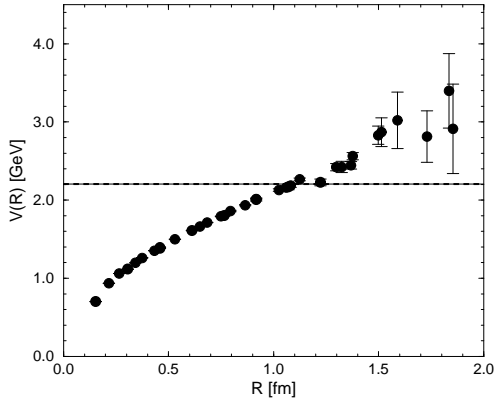}
\end{center}
\vspace{-18mm}
\caption{
The potential data on $16^3{\times}32$ lattice 
at $m_{\pi}/m_{\rho} \sim 0.6$.
The solid and dotted lines represent the center value of
$2m_{SL}$ and its error. }
\label{fig:potential}
\vspace{-19pt}
\end{figure}

In Fig.~\ref{fig:potential}
the potential data $V(R)$ are compared 
with twice the static-light meson mass $2 m_{SL}$
on the $16^3{\times}32$ lattice at the lightest sea quark mass.
The potential result crosses $2 m_{SL}$ at around 1~fm
and continues to increase with $R$. 
We find a similar behavior at other sea quark masses 
and lattice spacings. 
Thus our results do not show any indication
of a flattening of the potential 
expected from a pair creation of quarks, similar to those of 
previous studies\cite{stringbreaking.Lat97}.

A possible interpretation of this result is that 
the overlap of the Wilson loop operator 
with the state of a static-light meson pair is small.
In a simplified picture that the contribution of the meson pair to the 
Wilson loop is given by a static-light meson propagating along the 
loop, the overlap is given by a factor 
$\exp (-E_{SL}R)$ where $E_{SL}\!=\!m_{SL}\!-\!\delta m_{SL}$ with
$\delta m_{SL}\!\approx\! V_0/2$ is the binding energy 
of meson. 
Hence the contribution of the meson pair  
dominates over that of the string state 
only for sufficiently large $T$ satisfying  
$\exp (-2E_{SL} (R+T))\gg \exp ( -\sigma RT)$.
For numerical parameters appropriate for Fig.~\ref{fig:potential},
we find this condition to be met only 
for $T \gg 3$ fm at $R\approx1.5$fm, 
which is far too large for practical potential measurements. 
Devising an operator which does not suffer from the small overlap problem
is needed to observe the string breaking effect. 

\vspace{-1pt}
\section{Effect on the Coulomb coefficient}

\begin{figure}[tb]
\begin{center}
\vspace{-4mm}
\leavevmode
\epsfxsize=7.5cm
\epsfbox{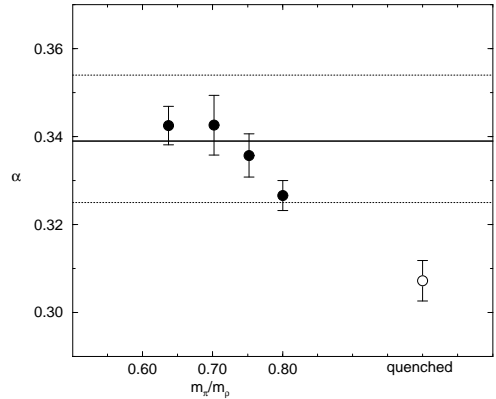}
\end{center}
\vspace{-18mm}
\caption{
Coulomb coefficient in full QCD(filled circle) as a function of 
$m_\pi/m_\rho$, 
as compared to the quenched value (open circle) 
at the same lattice spacing.
Solid and dashed lines represent an estimate expected for the full QCD 
value and error.}
\label{fig:alpha}
\vspace{-19pt}
\end{figure}

In Fig.~\ref{fig:alpha} we plot results for the Coulomb coefficient 
$\alpha$ obtained on the $24^3{\times}48$ lattice as a function of sea quark 
mass using $m_\pi/m_\rho$ (filled circles). The right-most open circle
is the result in quenched QCD at $\beta\!=\!2.7$ 
for which $a^{-1}_\sigma\!=\!2.217(8)$~GeV fixed 
by $\sqrt{\sigma}\!=\!0.44$~GeV  
matches with the value $a^{-1}_\sigma\!=\!2.22(4)$~GeV of the full QCD run.     
As expected, the full QCD results are larger than the quenched value, 
and they increase as the sea quark becomes lighter.

A simple estimate of the magnitude of shift of $\alpha$ from 
quenched to full QCD may be made in the following way\cite{shift.of.alpha}:
one starts with the quenched value, runs it from $\mu=a^{-1}$ 
down to $\mu\approx 500$ MeV using the 2-loop $\beta$ function 
with $N_f\!=\!0$, and then runs back to $a^{-1}$ with $N_f\!=\!2$. 
The result shown by the horizontal line in Fig.~\ref{fig:alpha} is 
consistent with the measured value. 
Dashed lines are 
uncertainties 
due to the error of the quenched value 
and the choice of the matching scale,
which is varied over $\mu=400$--$600$~MeV.  

%
%

\vspace{10pt}

We thank I. Drummond and B. Petersson for useful discussions. 
This work is supported in part by the Grants-in-Aid
of Ministry of Education
(Nos. 08640404, 09304029, 10640246, 10640248, 10740107).
GB, SE and KN are JSPS Research Fellows. 
HPS is supported by JSPS Research for
Future Program.

\end{document}